\documentclass[reprint,aps,prb,superscriptaddress, longbibliography]{revtex4-2}

\usepackage{graphicx}
\usepackage{dcolumn}
\usepackage{bm}
\usepackage{siunitx}
\usepackage{color}
\usepackage{booktabs}
\usepackage{miller} 
\usepackage{mathtools, stmaryrd}

\definecolor{r}{rgb}{0.86,0.08,0.23}
\usepackage{hyperref}
\definecolor{blue_n}{rgb}{0.,0.3,0.5}

\hypersetup{
backref=true, 
pagebackref=true,
hyperindex=true, 
colorlinks=true, 
breaklinks=true, 
citecolor = blue_n, 
urlcolor= blue_n,
linkcolor= black, 
pdftitle={Hopping of the center-of-mass of single G centers in silicon-on-insulator}, 
pdfauthor={  Alrik~Durand, Yoann~Baron, P\'eter~Udvarhely, F\'elix~Cache, Krithika~V.~R., Tobias~Herzig, Mario~Khoury, S\'ebastien~Pezzagna, Jan~Meijer, Jean-Michel Hartmann, Shay Reboh, Marco~Abbarchi, Isabelle~Robert-Philip, Adam~Gali, Jean-Michel~G\'erard, Vincent~Jacques, Guillaume~Cassabois and Ana\"is~Dr\'eau } 
}

\begin{document}

    \title{Hopping of the center-of-mass of single G centers in silicon-on-insulator}

	\author{Alrik Durand}\thanks{These authors contributed equally to this work.}	 	\affiliation{Laboratoire Charles Coulomb, Universit\'e de Montpellier and CNRS, 34095 Montpellier, France} 
	\author{Yoann Baron}\thanks{These authors contributed equally to this work.}			\affiliation{Laboratoire Charles Coulomb, Universit\'e de Montpellier and CNRS, 34095 Montpellier, France} 
	\author{P\'eter~Udvarhelyi}	\affiliation{Wigner Research Centre for Physics, P.O. Box 49, H-1525 Budapest, Hungary}
		\affiliation{Department of Atomic Physics, Institute of Physics, Budapest University of Technology and Economics, M\H{u}egyetem rakpart 3., H-1111, Budapest, Hungary}
	\author{F\'elix Cache}		\affiliation{Laboratoire Charles Coulomb, Universit\'e de Montpellier and CNRS, 34095 Montpellier, France} 
	\author{Krithika V.R. }		\affiliation{Laboratoire Charles Coulomb, Universit\'e de Montpellier and CNRS, 34095 Montpellier, France} 
	\author{Tobias~Herzig}		\affiliation{Division of Applied Quantum Systems, Felix-Bloch Institute for Solid-State Physics, University Leipzig, Linn\'eestra\ss e 5, 04103 Leipzig, Germany} 
	\author{Mario~Khoury}	 	\affiliation{CNRS, Aix-Marseille Universit\'e, Centrale Marseille, IM2NP, UMR 7334, Campus de St. J\'er\^ome, 13397 Marseille, France} 
	\author{S\'ebastien~Pezzagna}	\affiliation{Division of Applied Quantum Systems, Felix-Bloch Institute for Solid-State Physics, University Leipzig, Linn\'eestra\ss e 5, 04103 Leipzig, Germany} 
	\author{Jan~Meijer}		\affiliation{Division of Applied Quantum Systems, Felix-Bloch Institute for Solid-State Physics, University Leipzig, Linn\'eestra\ss e 5, 04103 Leipzig, Germany} 
	\author{Jean-Michel~Hartmann}  \affiliation{Univ. Grenoble Alpes, CEA, LETI,  F-38000 Grenoble, France} 
	\author{Shay~Reboh}	\affiliation{Univ. Grenoble Alpes, CEA, LETI,  F-38000 Grenoble, France} 
	\author{Marco~Abbarchi}	 \affiliation{CNRS, Aix-Marseille Universit\'e, Centrale Marseille, IM2NP, UMR 7334, Campus de St. J\'er\^ome, 13397 Marseille, France} 
		\affiliation{Solnil, 95 Rue de la République, 13002 Marseille, France}
	\author{Isabelle~Robert-Philip}   \affiliation{Laboratoire Charles Coulomb, Universit\'e de Montpellier and CNRS, 34095 Montpellier, France} 
	\author{Adam~Gali}	\affiliation{Wigner Research Centre for Physics, P.O. Box 49, H-1525 Budapest, Hungary}
		\affiliation{Department of Atomic Physics, Institute of Physics, Budapest University of Technology and Economics, M\H{u}egyetem rakpart 3., H-1111, Budapest, Hungary}
		\affiliation{MTA-WFK Lend\"ulet “Momentum” Semiconductor Nanostructures Research Group, Budapest, Hungary}
	\author{Jean-Michel~G\'erard}	\affiliation{Univ. Grenoble Alpes, CEA, Grenoble INP, IRIG, PHELIQS, 38000 Grenoble, France} 
	\author{Vincent Jacques}	 \affiliation{Laboratoire Charles Coulomb, Universit\'e de Montpellier and CNRS, 34095 Montpellier, France} 
	\author{Guillaume Cassabois}	 \affiliation{Laboratoire Charles Coulomb, Universit\'e de Montpellier and CNRS, 34095 Montpellier, France} 
		\affiliation{Institut Universitaire de France, 75231 Paris, France.}
	\author{Ana\"is Dr\'eau}	\email{anais.dreau@umontpellier.fr}	 \affiliation{Laboratoire Charles Coulomb, Universit\'e de Montpellier and CNRS, 34095 Montpellier, France} \email{anais.dreau@umontpellier.fr}

 \begin{abstract}

Among the wealth of single fluorescent defects recently detected in silicon, the G center catches interest for its telecom single-photon emission that could be coupled to a metastable electron spin triplet. 
The G center is a unique defect where the standard Born-Oppenheimer approximation breaks down as one of its atoms can move between 6 lattice sites under optical excitation. 
The impact of this atomic reconfiguration on the photoluminescence properties of G centers is still largely unknown, especially in silicon-on-insulator (SOI) samples. 
Here, we investigate the displacement of the center-of-mass of the G center in silicon. 
We show that single G defects in SOI exhibit a multipolar emission and zero-phonon line fine structures with splittings up to $\sim 1$ meV, both indicating a motion of the defect central atom over time. 
Combining polarization and spectral analysis at the single-photon level, we evidence that the reconfiguration dynamics are drastically different from the one of the unperturbed G center in bulk silicon. 
The SOI structure freezes the delocalization of the G defect center-of-mass and as a result, enables to isolate linearly polarized optical lines. 
Under above-bandgap optical excitation, the central atom of G centers in SOI behaves as if it were in a 6-slot roulette wheel, randomly alternating between localized crystal sites at each optical cycle.
Comparative measurements in a bulk silicon sample and \textit{ab initio} calculations highlight that strain is likely the dominant perturbation impacting the G center geometry. 
These results shed light on the importance of the atomic reconfiguration dynamics to understand and control the photoluminescence properties of the G center in silicon. 
More generally, these findings emphasize the impact of strain fluctuations inherent to SOI wafers for future quantum integrated photonics applications based on color centers in silicon.

    \end{abstract}

   \maketitle

\section{Introduction}

Color centers in silicon have recently received renewed attention as they could offer a new building block for quantum technologies \cite{beaufils_optical_2018, chartrand_highly_2018, redjem_single_2020,  simmons_single_2020, hollenbach_engineering_2020,  zhang_material_2020, durand_broad_2021, bergeron_silicon-integrated_2020, higginbottom_optical_2022, higginbottom_memory_2023, deabreu_waveguide-integrated_2023, johnston_cavity-coupled_2023, lee_high-efficiency_2023, islam_cavity-enhanced_2024, udvarhelyi_identification_2021, baron_single_2022, hollenbach_wafer-scale_2022, komza_indistinguishable_2022, lefaucher_cavity-enhanced_2023, prabhu_individually_2023, saggio_cavity-enhanced_2023, day_electrical_2023, durand_genuine_2024, berkman_millisecond_2023, weiss_erbium_2021, gritsch_purcell_2023, redjem_all-silicon_2023, zhiyenbayev_scalable_2023, baron_detection_2022, lefaucher_purcell_2024, udvarhelyi_l-band_2022}. 
Within a few years, single photon emission has been demonstrated on a dozen of different types of fluorescent defects in silicon \cite{redjem_single_2020, hollenbach_engineering_2020, durand_broad_2021, baron_detection_2022, higginbottom_optical_2022, baron_single_2022, hollenbach_wafer-scale_2022, gritsch_purcell_2023}. 
One of the next challenges is to provide single spin qubits interfaced with telecom light and integrated into silicon-on-insulator (SOI) photonic nanostructures \cite{higginbottom_optical_2022}. 
Therefore, strong emphasis has been placed on color centers with non-zero electron spin that could be optically initialized and readout, such as the T center \cite{bergeron_silicon-integrated_2020, higginbottom_optical_2022, higginbottom_memory_2023, deabreu_waveguide-integrated_2023, johnston_cavity-coupled_2023, lee_high-efficiency_2023, islam_cavity-enhanced_2024}, erbium dopants \cite{berkman_millisecond_2023, weiss_erbium_2021, gritsch_purcell_2023} or the G center \cite{lee_optical_1982, beaufils_optical_2018,  chartrand_highly_2018, udvarhelyi_identification_2021, baron_single_2022, hollenbach_wafer-scale_2022, komza_indistinguishable_2022, lefaucher_cavity-enhanced_2023, prabhu_individually_2023, saggio_cavity-enhanced_2023, day_electrical_2023, durand_genuine_2024}.  
This last defect is the brightest among the three and it can be isolated at single-defect scale in unstructured SOI wafers \cite{baron_single_2022}.

The G center in silicon is a fluorescent carbon-based complex studied for several decades \cite{foy_uniaxial_1981, thonke_new_1981, davies_carbon-related_1983}, and commonly used as a non-destructive probe of the carbon concentration \cite{davies_optical_1989}. 
Besides an optical emission in the telecom O-band, the G center features a metastable electron spin triplet whose magnetic resonances have been measured optically on ensembles \cite{lee_optical_1982, odonnell_origin_1983}. 
Single isolated G centers can be created by co-implantation of carbon ions and protons \cite{baron_single_2022} or by Si-focused ion beam implantation in carbon-implanted silicon with a spatial resolution of 50 nm \cite{hollenbach_wafer-scale_2022}. 
Purcell-enhanced emission has recently been demonstrated for ensembles of G centers \cite{lefaucher_cavity-enhanced_2023} and single G centers \cite{saggio_cavity-enhanced_2023} in photonic cavity, as well as Hong-Ou-Mandel interferences using an individual G defect integrated in a silicon waveguide \cite{komza_indistinguishable_2022}. 

Despite these advances, our knowledge on the photophysics of the G center in silicon remains limited. 
Unlike usual solid-state systems, this defect cannot be described under the standard Born-Oppenheimer approximation since its microscopic structure allows its center-of-mass to move \cite{udvarhelyi_identification_2021, lee_optical_1982, davies_carbon-related_1983,  davies_optical_1989, capaz_theory_1998}. 
How this structural reconfiguration impacts the optical properties of the G center in SOI is very little known, which hinders the development of potential quantum applications.

    \begin{figure}[h!]
        \includegraphics[width=0.88\columnwidth]{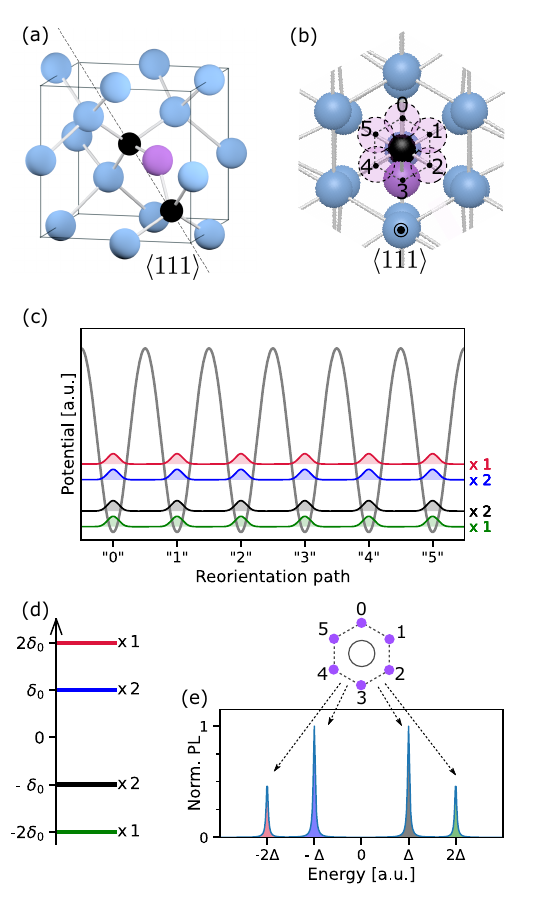}
        \caption{The G center in silicon in the unperturbed case. 
        (a) Microscopic structure of the defect. 
        Two substitutional carbon atoms C (\textit{black}) connected by an interstitial silicon atom Si$_{\mathrm{(i)}}$(\textit{purple}) are aligned along the $\langle 111 \rangle$ crystal axis. 
        (b) In the $\langle 111 \rangle$ direction, the G~center structure showing the 6 possible equivalent sites for the Si$_{\mathrm{(i)}}$. 
        (c) In gray, sketch of the energy potential of the 6 identical coupled wells experienced by the Si$_{\mathrm{(i)}}$ along its reorientation path around the C-C axis. 
        This potential is similar in both the electronic ground and excited states of the G center, albeit with different barrier heights \cite{udvarhelyi_identification_2021}. 
         In colors, density distributions for the 6 rotational states of the defect center-of-mass, with level degeneracy indicated on the right. 
         The rotational energy splittings are shifted for clarity and should not be compared to the potential barrier energy. 
        (d) Quartet energy level structure of the 6 delocalized states of the center-of-mass of the G center. 
        Energies are offset by $E_0$ (see main text). 
        (e) Theoretical zero-phonon line spectrum of the unperturbed G center in both absorption and photoluminescence.  
        $\Delta$ is the difference between the quartet energy splittings in the ground and excited states. 
        Each emission line corresponds to optical transitions between rotational states where the Si$_{\mathrm{(i)}}$ is equally delocalized over the 6 sites.  }
        \label{fig:unperturbed_G}
    \end{figure}

In this work, we explore the motion of the center-of-mass of the G center in silicon. 
We start by introducing the atomic configuration of the unperturbed defect and the energy level fine structure resulting from the rotational delocalization of its center-of-mass. 
Single G centers in SOI are then investigated through low-temperature photoluminescence (PL) experiments. 
Their single-photon polarization is first analyzed to reveal multipolar emission. 
We then examine the zero-phonon line spectra of these individual defects and show a fine structure with line splittings exceeding by two orders of magnitude the ones of unperturbed G centers. 
By combining spectral and polarization analysis, we demonstrate that the center-of-mass of these G centers is hopping over time between localized positions, like a 6-slot roulette wheel. 
Finally, performing \textit{ab initio} calculations and spectral measurements in a bulk silicon sample enables to evidence strain effects as the dominant perturbation freezing the rotational delocalization of single G centers in SOI structures. 
In conclusion, we discuss the impact of these findings on the control of the quantum properties of single G centers in SOI platforms.

\section{Unperturbed G center in silicon}
\label{sec:unperturbed}

The microscopic structure of the luminescent form of the G center in silicon was proposed in the 1980s \cite{odonnell_origin_1983, davies_carbon-related_1983, song_bistable_1990} and recently confirmed by advanced \textit{ab initio} calculations \cite{udvarhelyi_identification_2021}. 
This atomic configuration is displayed in Figure \ref{fig:unperturbed_G}(a). 
It consists of 2 substitutional carbon atoms aligned along the $\langle 111 \rangle$ crystal direction and connected by an interstitial silicon atom Si$_{\mathrm{(i)}}$. 
This defect configuration corresponds to a $C_{1h}$ symmetry, with a plane of symmetry defined by these 3 atoms \cite{udvarhelyi_identification_2021}. 
Unlike the tetravalent carbon atoms, the Si$_{\mathrm{(i)}}$ forms only two covalent bonds with its neighboring atoms and is therefore less rigidly attached to the matrix.
Indeed, recent density functional theory (DFT) simulations have shown that this silicon atom does rotate around the C-C axis \cite{udvarhelyi_identification_2021}. 
More precisely, in both ground and excited states, the interstitial silicon atom is delocalized by tunnel effect between the 6 equivalent sites indicated in Figure \ref{fig:unperturbed_G}(b) \cite{udvarhelyi_identification_2021}. 

Due to the displacement of its center-of-mass, the standard Born-Oppenheimer approximation breaks down for the G center in silicon. 
Nevertheless, this rotational degree of freedom can be introduced in a simple tight-binding model to extract the defect eigenstates and associated energies, as proposed in Ref. \cite{udvarhelyi_identification_2021}. 
Both in ground and excited states, the electronic wavefunction of the G defect can be written as the product of its center-of-mass wavefunction $\varphi (\bm{R}_{_{CM}})$ by the wavefunction of the electron cloud relative to the center-of-mass $\phi (\bm{r}-\bm{R}_{_{CM}})$:
\begin{equation}
\psi  (\bm{r}) = \varphi (\bm{R}_{_{CM}}) \cdot \phi (\bm{r}-\bm{R}_{_{CM}}) .
\end{equation}
To simplify notations, we omit the indices ${GS}$ and ${ES}$ referring to the ground and excited states, respectively.  
The DFT calculated potential energy surface (PES) along the reorientation path of the Si$_{\mathrm{(i)}}$ can be approximated by a 6-period cosine function in both ground and excited states, albeit with different barrier amplitudes \cite{udvarhelyi_identification_2021}. 
Due to the tunnel coupling between adjacent sites, the silicon atom thus behaves like a quasiparticle in a potential of 6 identical coupled wells (Fig. \ref{fig:unperturbed_G}(c)). 
Using the Bloch theorem, the 6 rotational eigenstates of the defect center-of-mass can be expressed in Dirac notation as:
\begin{equation}
| \varphi ^{(m)} \rangle = \frac{1}{\sqrt{6}} \sum_{n = 0}^{5} e^{ik_{m} n a} |n \rangle ,
\label{eq:phi_m}
\end{equation}
where $ m \in \llbracket 0,5 \rrbracket$, $k_m$ is the wavevector of the Bloch function, $a$ is the period of the potential and $| n \rangle$ the wavefunction describing the Si$_{\mathrm{(i)}}$ atom localized in the site $n$. 
From this equation, we directly get that the Si$_{\mathrm{(i)}}$ is equally delocalized on the 6 possible sites for each eigenstate of the G center (Fig. \ref{fig:unperturbed_G}(c)). 
The wavevector expression is derived from the Born-von Karman boundary conditions ($e^{ik_m 6 a} = 1$):
\begin{equation}
k_m = \frac{m \pi}{3 a}. 
\label{eq:k_m}
\end{equation}
The time-independent Schr\"odinger equation describing the stationary states of the defect center-of-mass is $\mathcal{H} | \varphi ^{(m)} \rangle = E^{(m)} | \varphi ^{(m)} \rangle$. 
Considering $E_0 = \langle n | \mathcal{H}  | n \rangle$ the energy of each uncoupled site, and  $\delta_0 = \langle n | \mathcal{H}  | n \pm 1 \rangle$ the tunneling coupling energy between adjacent sites, the energy of the eigenstate $|\varphi ^{(m)} \rangle$ writes as: 
\begin{equation}
E^{(m)} = E_0 + 2 \delta_0 \cos{(k_m a)}. 
\label{eq:E_m0}
\end{equation}
Combining Eqs. (\ref{eq:E_m0}) and (\ref{eq:k_m}), we finally obtain the formula for the 6 eigen-energies of the defect center-of-mass:
 \begin{equation}
E^{(m)} = E_0 + 2 \delta_0 \cos{\left( m \frac{\pi}{3} \right)}. 
\end{equation}
The rotational energy levels of the G center, in both ground and excited states, are thus arranged in a quartet fine structure with degeneracy 1:2:2:1 and splittings $\delta_0$-$2 \delta_0$-$\delta_0$, 
as represented in Figure \ref{fig:unperturbed_G}(d). 
However, due to a higher energy barrier for the Si$_{\mathrm{(i)}}$ rotation, the tunneling coupling energy is much smaller in the ground state \cite{udvarhelyi_identification_2021}. 
As a result, the quartet energy splitting in this level is negligible compared to the excited one: $ \delta_{0,_{GS}} \ll \delta_{0,_{ES}}$. 
We note that the same quartet structure is found for the electronic states of the benzene molecule \cite{udvarhelyi_identification_2021}.

In the Franck-Condon approximation, the selection rules for optical dipolar transitions between these levels are given by: 
\begin{equation}
\langle \psi_{GS} | - \bm{d}\cdot \bm{E}| \psi_{ES} \rangle = \langle \varphi_{GS}^{(m)} | \varphi_{ES}^{(m')}\rangle \cdot \langle \phi_{GS} | -\bm{d} \cdot \bm{E} | \phi_{ES} \rangle .
\label{eq:selection_rules}
\end{equation}
Since $\langle \varphi_{GS}^{(m)} | \varphi_{ES}^{(m')}\rangle= \delta_{m, m'}$ the Kronecker symbol, the rotational quantum number $m$ must be preserved during the optical transition. 
Hence optical transitions are only allowed between the same rotational eigenstates. 
Let $\Delta$ be the difference between the tunneling coupling energy in the ground and excited states: $\Delta = |\delta_{0,_{ES}} - \delta_{0, _{GS}}|$. 
From the optical selection rules, it results that a quartet fine structure with relative line intensities 1:2:2:1 and splittings  $\Delta$-$2 \Delta$-$\Delta$ should be observed in absorption and emission (Fig. \ref{fig:unperturbed_G}(e)). 
Indeed, such quartet spectra have been measured on an ensemble of G centers in isotopically purified bulk $^{28}$Si \cite{chartrand_highly_2018}. 
In both absorption and PL spectra, the energy splitting is $\Delta^{(exp)} = 2.5 \pm 0.2\, \mu$eV$= 0.60 \pm 0.05$ GHz. 
However, neither this quartet structure nor any other fine structure has been reported for G centers at single-defect scale. 

    \begin{figure}
        \includegraphics[width=0.9\columnwidth]{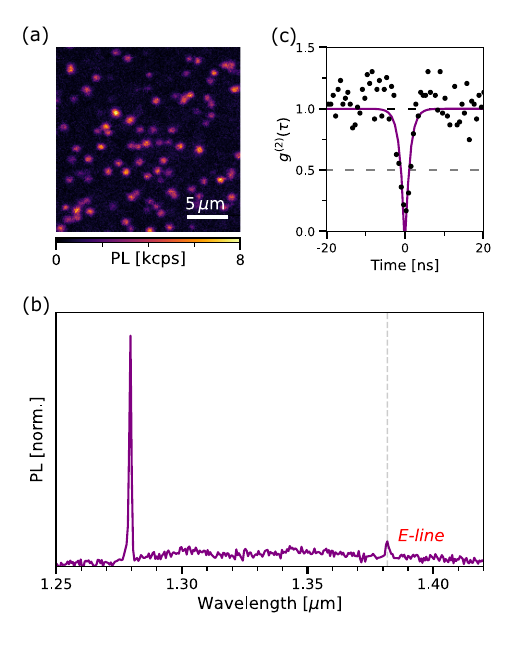}
        \caption{The G center in silicon.
        (a) Optical scan at 30 K of the SOI sample co-implanted with carbon ions and protons \cite{baron_single_2022}. 
        Hotspots are isolated G centers.
        (b) Low-resolution PL spectrum of a single G center showing the zero-phonon line at 1279 nm and the E-line associated with the defect local-vibrational mode.   
        (c) Second-order autocorrelation function demonstrating single-photon emission with an antibunching at zero delay $g^{(2)}(0) \simeq 0.17 < 1/2$.}
        \label{fig:intro}
    \end{figure}

             \begin{figure*}
        \includegraphics[width=0.9\textwidth]{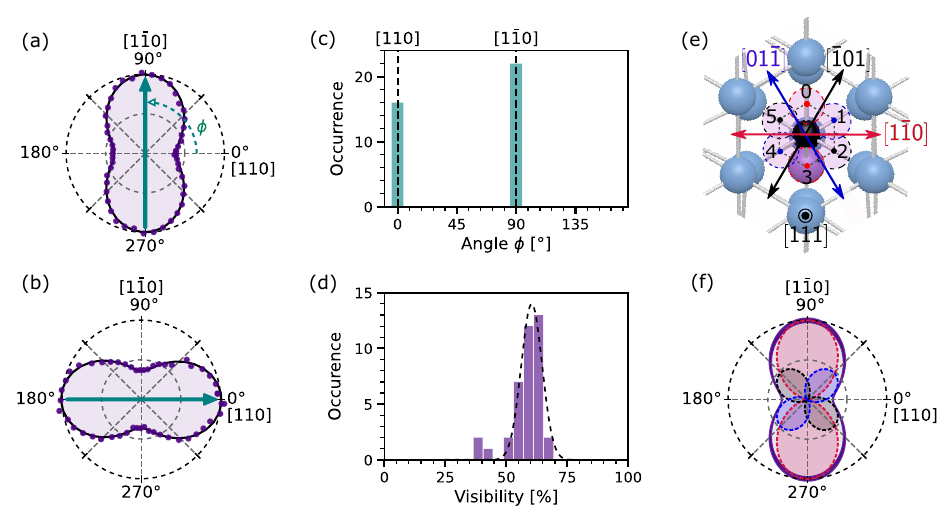}
        \caption{Polarization emission diagrams of single G centers in silicon. 
	  (a, b) Typical polarization emission diagrams recorded on two single G centers and corrected from background counts (see main text). 
	The diagram visibility $V$ and orientation angle $\phi$ are extracted from data fitting (\textit{solid lines}) with a function $V\cdot\cos^2(\theta-\phi) +1-V$. 
        (c, d) Histograms of the distribution of the angle $\phi$ and visibility $V$ measured on 39 {G~centers}, respectively. 
        (e) Orientations of the emission dipoles of a [111]-oriented G center depending on the position of the Si$_{\mathrm{(i)}}$: dipoles along $[1\bar{1}0]$,  $[01\bar{1}]$ and $[\bar{1}01]$ for the Si$_{(i)}$ positions $\{0,3\}$, $\{1,4\}$ and $\{2,5\}$, respectively. 
        (f) Polarization emission diagrams (\textit{dashed lines}) for each of the 3 dipoles considering photon collection from the (001)-top surface. 
        Their amplitudes correspond to the relative ratio of PL collected intensity (see main text). 
        The indigo thick line represents the total polarization diagram resulting from the contribution of the three dipoles (Appendix \ref{app:dip_eff}).
               }
        \label{fig:dipole}
    \end{figure*}

\section{Single G centers in SOI}
\label{sec:sample}

The silicon sample used here is the SOI wafer from Ref. \cite{baron_single_2022}. 
Its top silicon layer is made of 54 nm of isotopically pure $^{28}$Si grown at 650$^{\circ}$, 20 Torr with $^{28}$SiH$_4$ in a reduced pressure-chemical vapor deposition reactor, on top of 4 nm of silicon with natural isotope abundance. 
To create G centers, it was locally co-implanted with carbon ions and protons at 8keV and 6keV energies, respectively, using several fluences $\geq 3\cdot 10^{10}$ cm$^{-2}$ \cite{baron_single_2022}.
The experimental setup is a home-made confocal microscope built in a He-closed cycle cryostat, equipped with superconducting single-photon detectors (more details in Ref. \cite{redjem_single_2020, durand_genuine_2024}). 
We perform above-bandgap optical excitation using a 532-nm continuous laser and select sample areas in which well separated single G centers can be optically addressed. 
For practical reasons, all measurements have been performed at 30 K.

Optical raster scans of the SOI silicon sample with $^{28}$Si on top show well-isolated light spots (Fig. \ref{fig:intro}(a)). 
Their PL spectra enable an unambiguous identification of genuine G centers due to the presence of a zero-phonon line (ZPL) at 1278 nm and the E-line at 1382 nm (Fig. \ref{fig:intro}(b)) \cite{durand_genuine_2024}.
The single photon emission is confirmed by the observation of antibunching at zero delay in the autocorrelation function $g^{(2)}(\tau)$ of the PL signal (Fig. \ref{fig:intro}(c)).

        \begin{figure*}
        \includegraphics[width=0.85\textwidth]{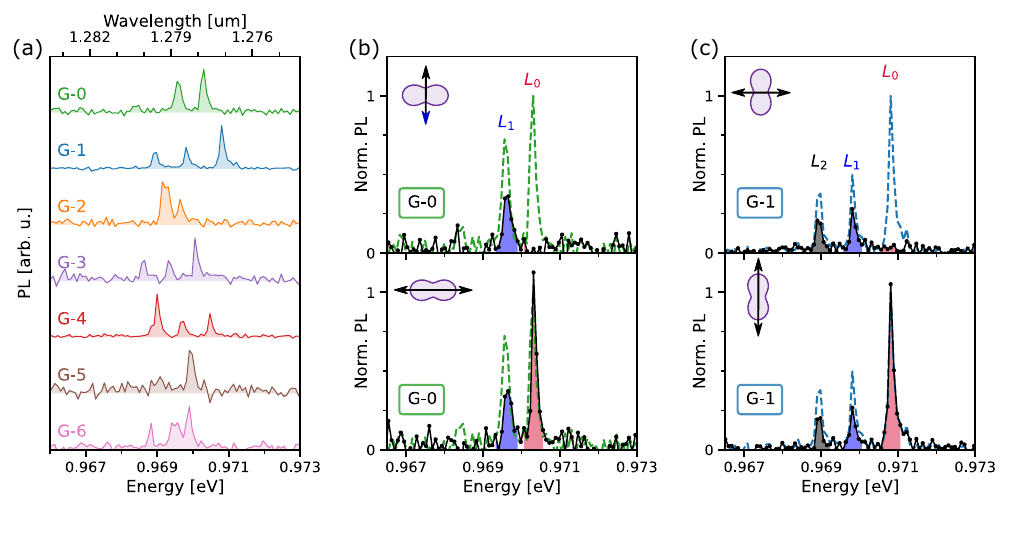}
        \caption{ Fine structure of single G centers in SOI. 
          (a) High-resolution ZPL spectra recorded on single G centers. 
       (b)[(c)] From top to bottom, ZPL spectrum measured on the defect G-0 [G-1] with a polariser selecting the polarization perpendicular and parallel to the main axis of the defect polarization diagram respectively, as indicated in the top left corners. 
       Note that the orientation of these 2 defects in the crystal is different: $[\bar{1}11]$ or $[1\bar{1}1]$ for G-0, and $[111]$ or $[\bar{1}\bar{1}1]$ for G-1. 
              The spectra with no polarization filtering from (a) are reproduced in dashed lines; the other PL spectra are normalized to its maximum.
              Colored areas represent the 0.3-meV integration range used in Figure \ref{fig:ZPL_theory}(a,b).  
        }
        \label{fig:ZPL_spectra}
    \end{figure*}

\section{Analysis of the single-photon polarization}
\label{sec:polar}

\subsection{Measurements}

Analysis of the polarization of single photons from G centers reveals a luminescence mechanism more complex than single dipole emission.
Emission polarization diagrams are acquired on single G defects by rotating the polarizer axis in the detection path.
PL photons are collected perpendicularly to the (001) top surface of the $^{28}$SOI sample.
To retain only the G center PL, diagrams are corrected from unpolarized background contribution measured at a single point close to defects. 
The first observation from Figure \ref{fig:dipole}(a-b) is that the polarization emission diagrams of single G centers are strongly anisotropic. 
For some G defects, the PL signal is maximum at 90$^{\circ}$, corresponding to the crystal axis $[1\bar{1}0]$ (Fig.~\ref{fig:dipole}(a)). 
In contrast for others, a PL maximum is detected at 0$^{\circ}$, i.e. towards the $[110]$ axis (Fig.~\ref{fig:dipole}(b)). 
A statistical analysis of 39 G centers in $^{28}$SOI shows that only these two directions are permitted, with 50/50 probability within statistical fluctuations (Fig.~\ref{fig:dipole}(c)). 
Another information from these diagrams is that no polarizer angle can cancel the PL signal from single G centers, as all diagrams show a visibility $\leq 68\%$ (Fig. \ref{fig:dipole}(d)). 
The average visibility over the set is 58\%, with a standard deviation of 7\%. 
The polarization diagrams of single G centers cannot be explained by the presence of a single emission dipole. 
As a consequence, the G center luminescence originates from multiple dipoles.

\subsection{Interpretation}

To understand the multipolar emission of the G center, we need to return to its microscopic structure. 
The interstitial silicon atom can take six positions in the plane perpendicular to the $\langle 111 \rangle$ defect main axis, as shown in Figure \ref{fig:unperturbed_G}(b). 
For a given static position of Si$_{(i)}$, symmetry point group theory for $C_{1h}$ symmetry authorizes a single emission dipole perpendicular to the symmetry plane \cite{udvarhelyi_identification_2021}. 
For a $[111]$-oriented G center, this leads to 3 pairs of possible dipoles aligned along the $[1\bar{1}0]$, $[01\bar{1}]$ and $[\bar{1}01]$ axes for the $\{0,3\}$, $\{1,4\}$ and $\{2,5\}$ Si$_{(i)}$ positions, respectively (Fig. \ref{fig:dipole}(e)). 
When projected onto the (001) sample top surface from which PL is collected, these dipoles produce polarization diagrams oriented at angles of $90^{\circ}$, $45^{\circ}$ and $135^{\circ}$ respectively, each with a visibility of 100\%. 
None of these dipoles taken alone can explain the single photon polarization observed from individual G centers (see Fig. \ref{fig:dipole}(a-d)). 
Numerical simulations with the finite-difference time-domain (FDTD) method show that the PL collected for the $[1\bar{1}0]$-dipole, that is parallel to the (001) top surface, is between 2.03 and 2.17 times more intense than that of the other two dipoles, depending on the defect depth in the Si layer (Appendix \ref{app:dip_eff}). 
Considering these three emission dipoles as independent and equally probable, we obtain the polarization diagram of Figure \ref{fig:dipole}(f), pointing towards $[1\bar{1}0]$ and with a visibility varying between 67.0\% and 68.5\% depending on depth.
The same polarization diagram would also be obtained for a $[\bar{1}\bar{1}1]$-oriented G center, while G centers aligned along $[\bar{1}11]$ or $[1\bar{1}1]$ would give diagrams turned by $90^{\circ}$.  
This model is in very good agreement with the experimental data of Figures \ref{fig:dipole}(a-d), allowing to conclude that the interstitial silicon atom of these single G centers does move over time during the measurements. 
Note that at this stage, it is not possible to deduce whether or not these G defects are in the unperturbed geometry described in Section \ref{sec:unperturbed}. 
Since the electron density is equally distributed over the 6 sites for all energy levels, each of the 3 dipoles has the same probability to emit and the polarization diagram should be identical to the one from Figure \ref{fig:dipole}(f). 
To gain more information about their center-of-mass rotation, we need to examine the fine structure of these individual G centers.

\section{ZPL Fine structure}
\label{sec:zpl}

\subsection{Measurements}

High-resolution PL spectra recorded on individual G defects show a fine structure radically different from the one presented in Figure \ref{fig:unperturbed_G}(e). 
Figure \ref{fig:ZPL_spectra}(a) displays the ZPL spectra recorded on 7 individual G centers in $^{28}$SOI for which antibunching with $g^{(2)}(0) <1/2$ has been measured with no spectral filtering.  
Except for one defect (G-5), all G centers show at least two or three ZPL lines.
Since single defects are investigated here, these multiple lines are evidence of a fine structure of the G center. 
This effect is to be distinguished from a lifting of orientational degeneracy under stress between non-equivalent defect orientations in ensemble of G centers \cite{tkachev_piezospectroscopic_1978, foy_uniaxial_1981, ristori_strain_2024}. 
These fine structures cannot be explained by a static configuration with no Si$_{\mathrm{(i)}}$ displacement, as the associated $C_{1h}$ monoclinic point group symmetry would only allow a single transition between $A'$ and $A''$ singlet orbital levels \cite{udvarhelyi_identification_2021}. 
The disparity in pattern and line positions observed between defects indicates an extrinsic origin, resulting from different local environments. 
Importantly, the quartet fine structure reported on unperturbed G centers has a characteristic splitting of 10 $\mu$eV, more than two orders of magnitude smaller than in Figure \ref{fig:ZPL_spectra}(a) \cite{chartrand_highly_2018}.  
Due to the spectral resolution of our spectrometer ($\simeq 150 \mu$eV), such quartet would appear as a single emission line in the spectra. 
The fine structures with meV-line splittings in Figure \ref{fig:ZPL_spectra}(a) rather indicate a significant perturbation of single G centers induced by their local environment.

    \begin{figure}
        \includegraphics[width=0.95\columnwidth]{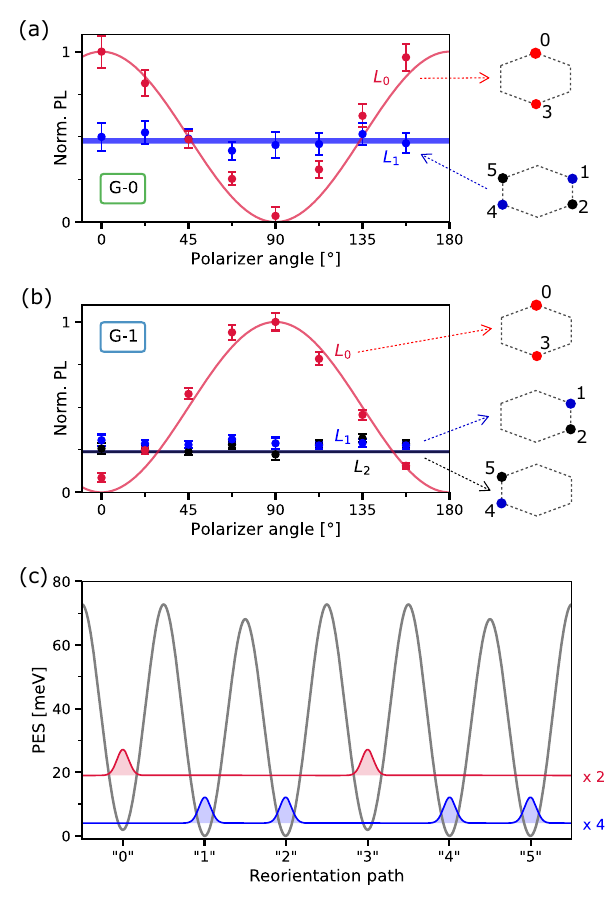}
        \caption{Signature of the localization of Si$_{\mathrm{(i)}}$ in single G centers. 
               (a, b) Evolution vs polarizer angle of the ZPL intensity for defects G-0 and G-1, respectively (see Fig.\ref{fig:ZPL_spectra}(b, c)).
               Solid lines are models considering the emission lines result from the Si$_{\mathrm{(i)}}$ sites indicated in the hexagons on the right.  	
               (c) In gray, calculated potential energy surface (PES) along the reorientation path of the interstitial silicon atom for a G center under  0.1\% $[1\bar{1}0]$-uniaxial strain. 
               In colors, numerical solutions for the resulting rotational states with level degeneracy indicated on the right. 
               The rotational energy splitting is increased for clarity and should not be compared to the energy scale on the PES.
               }
        \label{fig:ZPL_theory}
    \end{figure}

PL spectra of individual G centers in SOI strongly change with polarization filtering. 
We focus on the single G-0 and G-1 defects that have both well-resolved fine structures. 
The G-0 center shows a doublet fine structure with an energy separation of $0.70 \pm 0.02$~meV. 
As shown on Figure \ref{fig:ZPL_spectra}(b), when the polarizer is perpendicular to the main axis of the defect emission diagram, the highest-energy line $L_0$ is completely extinguished. 
When the polarizer is parallel, the $L_0$ line is similar to the previous spectrum with no polarization filtering. 
On the contrary, the second line $L_1$ has the same intensity for both polarizer angles, corresponding to roughly one half compared to the unfiltered spectrum.  
Similar behavior is observed on the triplet fine structure of the G-1 defect whose splittings from $L_0$ to $L_2$ are $1.00 \pm 0.02$~meV and $0.86\pm 0.02$~meV (Fig. \ref{fig:ZPL_spectra}(c)). 
The strongest line $L_0$ is globally unchanged for the polarizer parallel to the orientation of the G-1 diagram and cancels out for the perpendicular position. 
The intensity of the other two lines $L_1$ and $L_2$ is roughly decreased by a factor of 2 compared to the unfiltered spectrum in both cases.

The intensity of the different ZPLs is measured as a function of the polarizer angle for G-0 and G-1 centers in Figure \ref{fig:ZPL_theory}(a,b).  
For both centers, we observe that the $L_0$ line is linearly polarized as its intensity follows a sinusoidal modulation with a visibility close to unity: $90 \pm10\%$ and $94 \pm 5\%$  for G-0 and G-1, respectively. 
In contrast, the other lines are non-polarized because their intensities stay globally constant as a function of polarizer angle. 
Furthermore, the intensity ratios when the $L_0$ line is at maximum are roughly 2:1 and 4:1:1, for G-0 and G-1 respectively. 

\subsection{Interpretation}

The ZPL pattern evolution with polarization shows that the G defect environment lifts the quasi-degeneracy between the 3 emission dipoles of the unperturbed case. 
The  $L_0$ signal variation is not compatible with the multipolar emission described above and results instead from a single dipole. 
Since the $L_0$ intensity reaches a maximum at $0^{\circ}$ for G-0 [$90^{\circ}$ for G-1], we can associate it to the $[110]$-dipole [$[1\bar{1}0]$-dipole for G-1]. 
According to Figure \ref{fig:dipole}(f), such an emission can only take place if the Si$_{\mathrm{(i)}}$ is localized in one of the 2 interstitial sites $0$ or $3$ (Fig. \ref{fig:ZPL_theory}(a,b)).
To produce the $L_1$ and $L_2$ lines, the only possible explanation is an equal mixture of dipoles [101] and [011] for the center G-0 [dipoles $[\bar{1}01]$ and $[01\bar{1}]$ for G-1]. 
For the G-0 doublet, this implies that the $L_1$ line corresponds to the 4 positions $\{1,2,4,5\}$ of the interstitial silicon atom (Fig. \ref{fig:dipole}(e), \ref{fig:ZPL_theory}(a)). 
Assuming that all Si$_{\mathrm{(i)}}$ sites are equally occupied over the time of the acquisition, the relative intensity of $L_1$ after the polarizer should be $\simeq 1/2.1$ of the maximum intensity of $L_0$ according to FDTD simulations (Appendix \ref{app:dip_eff}). 
Within error bars, these values well match the experimental data for G-0 displaying maximal intensity ratio of 2:1, as shown on Figure \ref{fig:ZPL_theory}(a). 
Following the same reasoning for defect G-1, possible Si$_{\mathrm{(i)}}$ position configurations for the lines $L_2$ and $L_1$ would be $\{1,2\}$ \& $\{4,5\}$ or $\{1,5\}$ \& $\{2,4\}$ (Fig. \ref{fig:ZPL_theory}(b)). 
Furthermore, their relative intensity should be half that of the line $L_1$ of G-0 associated to twice more sites. 
This interpretation is in good agreement with the experimental data for G-1 showing maximal intensity ratios of roughly 4:1:1 (Fig.~\ref{fig:ZPL_theory}(b)). 
We note that the G-3 center that also has a triplet fine structure shows the same polarization behavior as G-1 (see Appendix \ref{app:G-3}). 
It is therefore very likely that these 2 defects experience a similar perturbation at first order. 

The ZPL evolution with polarization establishes that the defect center-of-mass is not in delocalized rotational states, as expected for unperturbed G centers (Sec. \ref{sec:unperturbed}). 
Instead, these G defects experience an external coupling distorting their original symmetry and freezing the rotational dynamics of their center-of-mass. 
Partial delocalization of Si$_{\mathrm{(i)}}$ could still happen between adjacent sites (i.e. for $\{1,2\}$ or $\{4,5\}$), but only if their energy difference is on the order of or smaller than the tunneling coupling energy.  
Given the energy scale at stake here compared to $\delta_{0,_{ES}}\simeq 2.5 \mu$eV (see Sec. \ref{sec:unperturbed}), we can consider this tunneling effect to be unlikely for the single G centers under study. 
Consequently, the center-of-mass eigenstates in both ground and excited states are the wavefunctions localized on each of the 6 Si$_{\mathrm{(i)}}$ sites, and not the Bloch states derived previously in Section \ref{sec:unperturbed}. 
The selection rules derived in the Franck-Condon approximation naturally (Eq.\ref{eq:selection_rules}) imply that the localized site of the defect center-of-mass is preserved during optical emission. 
As a result, the ZPL energy reflects the energy difference of one given Si$_{\mathrm{(i)}}$ site between the excited and ground states. 
Since the G center has comparable geometric configurations in these two states, the ZPL splitting pattern directly indicates the energy degeneracy lifting between the different Si$_{\mathrm{(i)}}$ sites.

Although the optical transitions preserve the configuration of the defect, our observations show that the Si$_{\mathrm{(i)}}$ of these single G centers does move over the acquisition time. 
More precisely, it spends on average the same amount of time in each of the 6 possible sites.
This motion cannot be thermally-activated hopping as $k_B T$ ($\simeq 2.6$ meV at 30K) is much smaller than the potential barrier energies in both the ground and excited states (89 and 33 meV, respectively from Ref. \cite{udvarhelyi_identification_2021}).
Instead the motion of the center-of-mass of G centers could result from the excess energy provided to the system during the above-bandgap optical excitation. 
Furthermore, this off-resonant excitation is not governed by the selection rules derived above for resonant optical transitions (Eq. \ref{eq:selection_rules}). 
At each optical excitation, the interstitial silicon of the G center randomly hops between the different positions, like a ball in a 6-slot casino roulette wheel. 
By repeating the optical cycles, each Si$_{\mathrm{(i)}}$ site is on average equally occupied, thus explaining our experimental data from Sections \ref{sec:polar},\ref{sec:zpl}. 
The next step is to determine which perturbation is responsible for blocking the delocalization dynamics of these individual G centers and generating the observed fine structures.

\section{Perturbations from SOI structures}
\label{sec:perturbations}

    \begin{figure}[h!]
        \includegraphics[width=0.95\columnwidth]{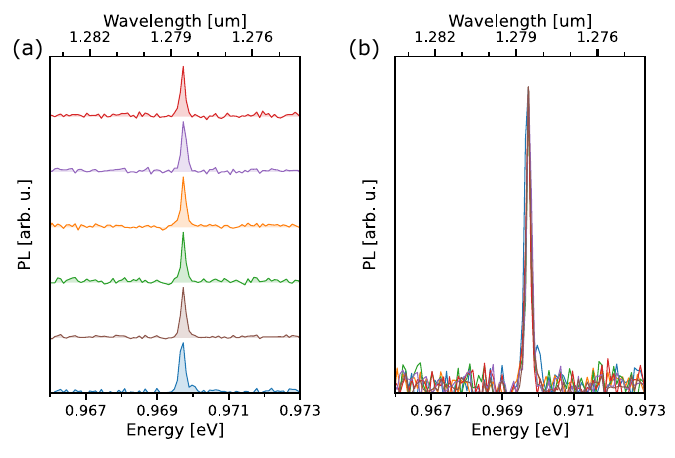}
        \caption{ZPL spectra recorded on isolated G center spots in a bulk $^{28}$Si sample plotted (a) with and (b) without vertical offset.  }
        \label{fig:ZPL_bulk}
    \end{figure}

Since single G centers are incorporated in a SOI structure, the first perturbation that comes to mind is lattice strain. 
Indeed, the difference in thermal expansion coefficients between Si and SiO$_2$ leads to in-plane strain in the top silicon layer \cite{camassel_strain_2000}. 
Furthermore, even state-of-the-art commercial SOI wafers are known to present lattice undulations at the $\mu$m-scale resulting from the bonding process used during fabrication \cite{fukuda_white_2006}. 
To test whether the ZPL splitting patterns come from the SOI structure, we first investigate G defects fabricated with a similar co-implantation method in a bulk silicon sample made of 1.5 $\mu$m of $^{28}$Si on top of a natural Si substrate. 
G defects are created using a fluence of $10^{10}$cm$^{-2}$ for both carbon atoms and protons, and with the same energies as for the SOI sample (Sec. \ref{sec:sample}). 
Figure \ref{fig:ZPL_bulk} shows the ZPL spectra measured on randomly chosen isolated G center spots in the bulk sample. 
For all investigated defects, a single ZPL line at 1278.6 nm was observed and without any ZPL fine structure found.
These results demonstrate that the frozen delocalization of the G defect center-of-mass is indeed due to its formation in the SOI structure.

To evaluate the impact of homogeneous uniaxial strain on the G center, we apply DFT calculations and map the PES of the interstitial silicon atom of the defect (see Appendix \ref{app:theory} for DFT methods). 
Since the G center has similar geometric structures in both ground and excited states, we can focus only on the external perturbation in the ground state. 
As optical transitions occur between the same localized Si$_{\mathrm{(i)}}$ sites, the splitting pattern in the ground state rotational levels qualitatively explains the fine structure of the ZPL.

The \textit{ab initio} PES calculated for a G center under  $0.1\%$ $[1\bar{1}0]$ uniaxial strain is displayed as a gray curve in Figure \ref{fig:ZPL_theory}(c). 
Unlike the unperturbed case from Figure \ref{fig:unperturbed_G}(c), the PES is not a sixfold symmetric periodic potential anymore.
In particular, the well minima of sites $0$ and $3$ are shifted by $\simeq 1.9$ meV compared to the 4 other sites (Fig. \ref{fig:ZPL_theory}(c)).   
Solving the time-independent Schrödinger equation for this perturbed potential leads to 2 separated subsets of rotational states, as shown on Figure \ref{fig:ZPL_theory}(c). 
The highest energy sublevel corresponds to the 2 degenerate states localized on sites $0$ and $3$, whereas the lowest one is associated to the 4 degenerate states localized on sites $\{1,2,4,5\}$. 
Consequently, such $[1\bar{1}0]$ uniaxial strain does not only freeze the delocalization of the Si$_{\mathrm{(i)}}$ of the G center but reproduces the doublet fine structure observed on G-0 defect (Fig. \ref{fig:ZPL_theory}(a)). 
Our calculations indicate that the origin of the ZPL splitting pattern is the energy difference of the center-of-mass potential minima between the excited and ground states. 
To explain the G center triplet fine structure, an extra perturbation needs to be added to the uniaxial strain to split the 4-fold sublevel (Fig.~\ref{fig:ZPL_theory}(b)).
A possible candidate could be electric field, possibly coming from localized charges at the silicon-to-oxide interface \cite{schroder_semiconductor_2005}. 
The impact of electric field on G centers is beyond the scope of the paper and will be investigated in future works.

At last, to compare different uniaxial strain directions, the ZPL energies are estimated using the Delta self-consistent-field ($\Delta$SCF) method for each set of equivalent equilibrium frozen configurations of the G center \cite{gali_theory_2009}.
For the 4 simulated strain directions, namely $[1\bar{1}0]$, $[110]$,  $[001]$ and $[111]$, only the magnitude of the ZPL splitting and the overall energy shift of the spectrum differ between different directions (see Appendix \ref{app:theory}). 
Especially, hardly no splitting is observed for the [111]-strain because this totally symmetric perturbation equally modifies the barrier energy for all sites at first order. 
Assuming a linear strain regime, this calculation method qualitatively reproduces the ZPL splittings observed on ensembles of G centers under uniaxial stress along $(001)$, $(111)$ and $(110)$ \cite{foy_uniaxial_1981} (see Appendix \ref{app:theory}). 
This confirms its validity for estimating the behavior of the G center ZPL doublet with strain direction.

\section{Conclusion}

In this work, we have investigated the center-of-mass motion of the G center in silicon resulting from the motion of its intrinsic interstitial silicon atom between 6 crystal sites. 
For the unperturbed G center, the distinctive quartet fine structure arising from the Si$_{\mathrm{(i)}}$ perfect delocalization by tunnel effect and the subsequent selection rules for the optical transitions can be derived from a simple tight-binding model. 
Single-photon polarization analysis from individual defects shows that the G~center emission emanates from 3 dipoles matching the 6 possible sites of the Si$_{\mathrm{(i)}}$. 
Individual G centers in $^{28}$SOI exhibit ZPL fine structures that do not only strongly vary from one center to another, but above all with $\sim1$-meV energy splittings. 
Such ZPL splittings are more than 2 orders of magnitude larger than the one observed on an ensemble of unperturbed G centers in bulk $^{28}$Si \cite{chartrand_highly_2018}. 
The evolution of the ZPL patterns with polarization evidences that the optical transitions of these single G centers are not compatible with a center-of-mass delocalization and instead occur while the Si$_{\mathrm{(i)}}$ sits at a fixed position, chosen among the 6 potential minima. 
\textit{Ab initio} simulations identify uniaxial strain as being the dominant mechanism blocking the G center tunneling rotation by 
partially lifting the energy degeneracy between the different Si$_{\mathrm{(i)}}$ sites. 
The correlation between the strain inherent to the SOI structure and the giant ZPL fine structure measured on single G centers is corroborated by the lack of any ZPL splitting for G centers fabricated with the same ion implantation method but in a $^{28}$Si bulk sample. 
As above-bandgap excitation does not obey the selection rules for direct optical transitions, we believe this off-resonant excitation causes the Si$_{\mathrm{(i)}}$ to hop randomly between the 6 sites, as if it were shaken in a 6-slot roulette wheel.

The current results are a milestone on the road to controlling the quantum properties of the G~center in silicon. 
The sensitivity of its fine structure to lattice distortion could be used as sub-micron scale optical strain sensor in SOI based nanomechanical resonators \cite{craighead_nanoelectromechanical_2000}. 
Furthermore, strain tuning of the defect optical emission wavelengths could be an alternative pathway to standard electrical tuning based on Stark effect \cite{bassett_electrical_2011, de_las_casas_stark_2017}. 
As the motion of the Si$_{\mathrm{(i)}}$ is also allowed in the metastable electron spin triplet level \cite{lee_optical_1982, odonnell_origin_1983, vlasenko_spin-dependent_1986, udvarhelyi_identification_2021}, investigating the rotational dynamics of the G center could also be performed through its spin quantum degree of freedom.

Our results highlight the key role on the properties of color centers of the inhomogeneous strain field in SOI films. 
The resulting ZPL fluctuations between defects, as shown here for G centers, will have to be efficiently compensated in view of developing quantum photonic chips relying on the emission from multiple color centers. 
However far from being only a drawback, strain perturbation looks as an attractive resource in the specific case of the G center as it enables to isolate single-dipole optical transitions following the freezing of its center-of-mass delocalization. 
This opens the route towards an efficient coupling to single-mode SOI nanophotonic cavities for cavity quantum electrodynamics experiments. 
For instance, in contrast, Purcell enhanced PL will be reduced for the unpolarized optical transitions of unperturbed G centers \cite{gerard_solid-state_2003}.
In this context, it is worth mentioning that suspended silicon nanobeams fabricated by selectively etching the buried silica are likely to remove the strain fluctuations resulting from the Si-SiO$_2$ interface \cite{fukuda_white_2006}.
Combined with uniaxial strain from tensor arms \cite{chretien_gesn_2019,guilloy_tensile_2015}, these nanostructures could provide a homogeneous strain environment for color centers.
As a result, the G centers will exhibit highly uniform optical emission lines with no center-of-mass delocalization. 
Taking benefit of these recent advances, one could use strain to engineer the PL properties of G centers in silicon in a deterministic way.

\section*{Acknowledgments}

We acknowledge funding from the French National Research Agency (ANR) through the projects OCTOPUS (No. ANR-18-CE47-0013-01) and QUASSIC (No. ANR-18-ERC2-0005-01), the Plan France 2030 through the project OQuLuS ANR-22-PETQ-0013, the Occitanie region through the SITEQ contract and the European Research Council (ERC) under the European Union’s Horizon 2020 research and innovation programme (project SILEQS, Grant No. 101042075). 
A. Durand acknowledges support from the French DGA.
A. Gali acknowledges the support  by the Ministry of Culture and Innovation and the National Research, Development and Innovation Office within the Quantum Information National Laboratory of Hungary (Grant No. 2022-2.1.1-NL-2022-00004). 
A. Gali additionally acknowledges the high-performance computational resources provided by KIFÜ (Governmental Agency for IT Development) institute of Hungary and the European Commission for project QuMicro (Grant No. 101046911).
The authors thank François Rieutord for enlightening discussions regarding lattice strain in SOI wafers. 

\appendix

\addcontentsline{toc}{section}{Appendices}
\renewcommand{\appendixname}{APPENDIX}
\renewcommand{\thesection}{\Alph{section}}

\section{\uppercase{Dipole collection efficiencies}}
\label{app:dip_eff}

      \begin{figure}[h!]
        \includegraphics[width=0.95\columnwidth]{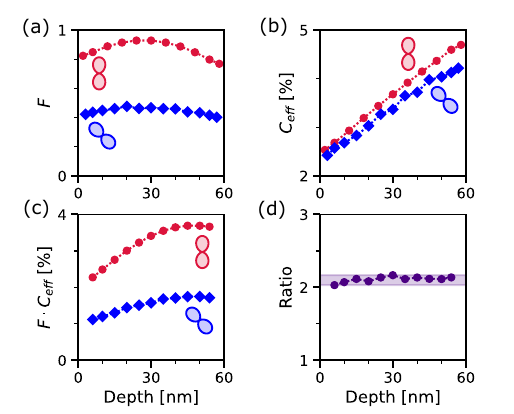}
        \caption{Simulations of the dipole collected intensity as a function of the defect depth. %
        (a-c) Respectively, Purcell factor~$F$, collection efficiency $C_{\mathrm{eff}}$ and their product for the dipoles $[1\bar{1}0]$ (red) and $[01\bar{1}]$ (blue). 
        (d) Evolution of the relative PL contribution between these two dipoles. }
        \label{fig:dip_eff}
    \end{figure}

The relative PL contribution used in Figure \ref{fig:dipole}(f) for the G center dipoles inside the SOI sample is simulated using the fine-difference time-domain method on the commercial software RSOFT from Synopsis (version 19.09). 
Since the top surface of the SOI sample is (001)-oriented, both $[\bar{1}01]$ and  $[01\bar{1}]$ dipoles will lead to the same results (see Fig.~\ref{fig:dipole}(e)).
The dipole spontaneous emission enhancement factor $F$ resulting from Purcell effect in the top silicon layer is computed as function of the defect depth for $[1\bar{1}0]$ and $[01\bar{1}]$ dipoles (Fig.~\ref{fig:dip_eff}(a)). 
The value of $F$ indicates whether the dipole emission is enhanced ($F\!>\!1$) or reduced ($F\!<\!1$) compared to its value in bulk silicon.
While the emission of the $[1\bar{1}0]$ in-plane dipole is little affected, the one of the $[01\bar{1}]$ out-of-plane dipole is inhibited by a factor of $\simeq 2$  due to the SOI structure. 
On the contrary, the dipole orientation has a small impact on its collection efficiency $C_{\mathrm{eff}}$ through a 0.85-Numerical Aperture microscope objective, as shown on Figure \ref{fig:dip_eff}(b). 
However, it strongly varies with the dipole depth, from $\simeq 2.5\%$ at the surface to $\simeq 4.5\%$ at the oxide interface. 
The total collected PL is proportional to the product $F\cdot C_{\mathrm{eff}}$ for each dipole (Fig.~\ref{fig:dip_eff}(c)). 
To obtain the relative contribution between the in-plane and out-of-plane dipoles, we compute the ratio $r= (F^{[1\bar{1}0]}\cdot C_{\mathrm{eff}}^{[1\bar{1}0]})/(F^{[01\bar{1}]}\cdot C_{\mathrm{eff}}^{[01\bar{1}]})$. 
As displayed in Figure \ref{fig:dip_eff}(d), the defect depth has little influence on this ratio that evolves in the [2.0-2.2] range within the 60-nm thick silicon layer.

\section{\uppercase{Triplet fine structure analysis on defect G-3}}
\label{app:G-3}

    \begin{figure}[h!]
        \includegraphics[width=0.9\columnwidth]{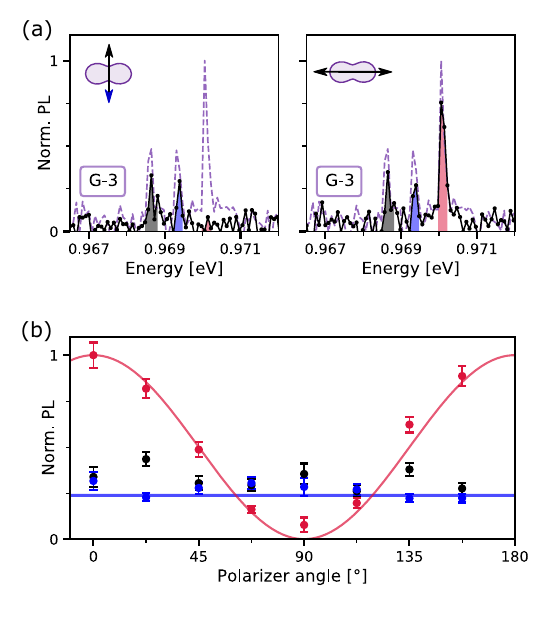}
        \caption{Triplet ZPL fine structure of G-3 defect. 
         (a) From left to right, ZPL spectrum measured on the defect G-3 with a polariser selecting the polarization perpendicular and parallel to the main axis of the defect polarization diagram respectively, as indicated in the top left corners. 
              The spectra with no polarization filtering from Figure \ref{fig:ZPL_spectra}(a) are reproduced in dashed lines; the y-axis is normalized to its maximum.
              Colored areas represent the 0.2-meV integration range used in (b). 
        (b) Evolution vs polarizer angle of the intensity of the 3 ZPL (see Fig.\ref{fig:ZPL_spectra}(b,c)).
              Solid lines are models considering the emission lines result from the Si$_{\mathrm{(i)}}$ sites indicated in figure \ref{fig:ZPL_spectra}(b).  }
        \label{fig:G-3}
    \end{figure}

The spectral analysis with polarization from Section \ref{sec:zpl} is performed on the single G center G-3, that presents a triplet ZPL fine structure. 
The evolution of its ZPL with polarization filtering reproduces the behavior observed on the triplet fine structure of defect G-1 (Fig.~\ref{fig:ZPL_spectra}(c), \ref{fig:ZPL_theory}(b)). 
Especially, we measure again that the highest optical line is fully cancelled for polarization perpendicular to the defect emission diagram (Fig.~\ref{fig:G-3}(a)).
The quantitative analysis of the ZPL intensities as function of polarizer angle shows that this line is indeed linearly polarized while the other 2 ZPL are unpolarized (Fig.~\ref{fig:G-3}(b)). 
Like the single G centers in SOI previously investigated, the PL from the G-3 defect results from optical transitions occuring between localized states of its center-of-mass. 
In particular, the ZPL fine structure of the G-3 center could be explained by the same localized Si$_{\mathrm{(i)}}$ sites as for the G-1 center, such as the ones proposed in Figure \ref{fig:ZPL_theory}(b).

\section{\uppercase{Ab-initio calculations for perturbed G centers}}
\label{app:theory}

\subsection{DFT methods}

 We apply DFT calculations on a G center perturbed by uniaxial strain and map the ground state PES along the rotational reorientation path of the central silicon atom of the G center. 
   Along the associated $Q$ vibrational coordinate, we sample all the 6 minima and barrier points and one additional middle point between these, using 24 single-point calculations in total for a single PES calculation. 
   These geometric configurations are relaxed by climbing image nudged elastic band (NEB) method on the sampling images until a $10^{-4}~\mathrm{eV/\AA}$ threshold is reached in the forces. This precision is necessary for the accurate modeling of small perturbations resulting in small relative energy differences. Consequently, we need to compromise the accuracy in the total energy by using PBE functional. 
   The calculations are performed in the plane wave based Vienna Ab initio Simulation Package (VASP)~\cite{blochl_projector_1994}. 
   The defect is modeled in a large 512-atom supercell with single $\Gamma$-point sampling in the reciprocal space. The core electrons are treated in the PAW formalism~\cite{kresse_ab_1993, kresse_efficient_1996, kresse_efficiency_1996, paier_screened_2006}. 
   The atomic positions remain fixed during this calculation. 
   The rotational reorientation is modeled in analogy to our previous work in Ref.~\cite{udvarhelyi_identification_2021}.

       \begin{figure}[h!]
        \includegraphics[width=0.9\columnwidth]{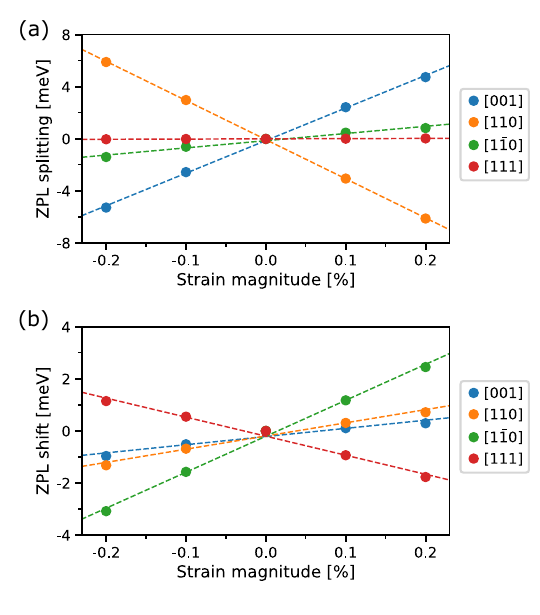}
        \caption{Energy of the (a) ZPL doublet splitting and (b) average ZPL shift for a [111]-oriented G center under various directions of uniaxial strain. }
        \label{fig:strains}
    \end{figure}

 \subsection{ZPL patterns for different uniaxial strains}

    \begin{figure*}
        \includegraphics[width=0.7\textwidth]{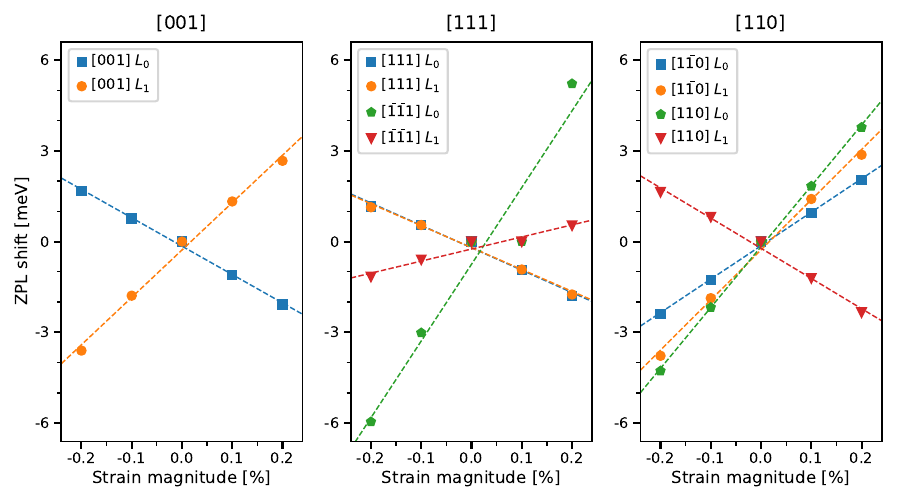}
        \caption{Evolution of the energy shift of the 2 ZPL $L_0$ and $L_1$ of a $[111]$-oriented G defect under different uniaxial strain directions to mimic the behavior of an ensemble of G centers under $[001]$, $[111]$ and $[110]$ uniaxial strain, respectively. }
        \label{fig:ensembles}
    \end{figure*}
 
We estimate the evolution of the ZPL patterns under different strain directions using the $\Delta$SCF method in the non-equivalent local minima positions as calculating the PES for each direction is very heavy on computational resources.
We can safely assume that the ZPL shift is dominated by the relative energy shift in the local minima of the PES for the larger strain regimes where frozen position approximation holds.
We note that this simulation method is not accurate to predict the correct ZPL energy ($\simeq 654$ meV for the unperturbed G center) but it enables to qualitatively extract strain-induced ZPL splitting trends, as shown in the following.
The ZPL energy is calculated for each subset of equivalent equilibrium Si$_{\mathrm{(i)}}$ positions for 4 uniaxial strain directions: $[001]$, $[110]$, $[1\bar{1}0]$ and $[111]$.
As visible on Figure~\ref{fig:strains}(a), the ZPL splitting energies are proportional to the strain amplitude in the $[-0.2\%, +0.2\%]$ investigated range. 
The largest splitting is obtained for the $[110]$-strain, while hardly no splitting is observed for the $[111]$-strain, as it corresponds to a totally symmetric perturbation. 
All 4 simulated strain directions produce an overall shift of the ZPL doublet fine structure (Fig.~\ref{fig:strains}(b)).

As a safety check, we calculate the evolution of the ZPL patterns of an ensemble of G centers subject to uniaxial strain along the crystal directions used in a previous piezospectroscopic experiment \cite{foy_uniaxial_1981} (Fig. \ref{fig:ensembles}). 
For a $[001]$-strain, all 4 possible $\langle 111 \rangle$ crystal axis of the G center experience the same lattice distortion, hence the same doublet fine structure. 
Under a $[111]$-strain, the $[111]$-oriented G centers will feel a totally symmetric perturbation with no splitting while the 3 other G orientations will undergo a perturbation equivalent to a $[\bar{1}\bar{1}1]$-strain for [111]-oriented G center. 
Consequently, 3 optical lines are expected to split for a G-center ensemble under [111]-strain. 
The $[110]$ strain will separate the orientations into 2 families: $[111]$- and $[\bar{1}\bar{1}1]$-oriented centers on one hand, and $[\bar{1}11]$ and $[1\bar{1}1]$-oriented defects on the other hand. 
The strain experienced by these last 2 orientations will be equivalent to a $[1\bar{1}0]$-strain acting on a [111]-oriented G center. 
As a result, 4 optical lines are expected to emerge for an ensemble of G defects under $[110]$-uniaxial strain. 
Assuming the linear regime of Hooke's law prevails, the ZPL splittings calculated for these 3 uniaxial strains are in good qualitative agreement with the uniaxial stress measurements performed on a ensemble of G centers in \cite{foy_uniaxial_1981}.

  \newpage

\bibliography{biblio_G_roulette}


\end{document}